\documentstyle[psfig,aps,twocolumn]{revtex}

\newif\ifels\elsfalse
\newif\iftt\ttfalse

\def\riga#1{}

\begin{document}
\ifpreprintsty\else
\twocolumn[\hsize\textwidth%
\columnwidth\hsize\csname@twocolumnfalse\endcsname
\fi
\ifels
{\noindent Date submitted: 6/7/1997\\[.5cm]}
\def\title#1{{\large\bf #1}\\[1cm]}
\def\author#1{#1\\[5mm]}
\def\address#1{{\em #1}\\[5mm]}
\def\maketitle{}
\def\date#1{}
\long\def\keywords#1{\vskip5mm
Keywords: #1
\vfill\noindent
RevTEX source available from\hskip5mm
\parbox[t]{0.6\textwidth}{\tighten
Sergio Conti\\
Scuola Normale Superiore\\
Piazza dei Cavalieri 7\\
I--56126 Pisa, Italy\\
Tel +39 50 509038, Fax +39 50 563513\\
e-mail: {\tt conti@ella.sns.it}}}
\else
\def\keywords#1{}
\fi
\title{\riga{20}{Bosonization theory for tunneling spectra in smooth
edges of Quantum Hall systems}}
\author{Sergio Conti$^\dagger$ and Giovanni Vignale$^\#$}
\address{\riga{65}{$^\dagger$ Scuola Normale Superiore, 56126 Pisa, Italy}}
\address{\riga{65}{$^\#$Department of Physics,  University
of Missouri, Columbia,  Missouri 65211, USA}}
\date{\today}
\maketitle
\begin{abstract}\riga{65}{
We calculate the spectral function of a smooth edge of a quantum Hall
system in the lowest Landau level by means of a bosonization
technique. We obtain a general relationship between the one electron
spectral function and the dynamical structure factor. The resulting
$I$--$V$ characteristics exhibit, at low voltage and temperature,
power law scaling, generally different from the one predicted by the
chiral Luttinger liquid theory, and in good agreement with recent
experimental results.}
\end{abstract}
%
\ifpreprintsty\clearpage\else\vskip1pc]\narrowtext\fi
\riga{40}
Edges of Quantum Hall systems have been studied mainly as realizations
of exotic correlated states of electrons. For example a sharp edge of
a fractional quantum Hall fluid is described as a Chiral Luttinger
Liquid (CLL)\cite{Wen90_91_92}. The main characteristic of a CLL
is the lack of single 
particle excitations at the Fermi level: the elementary excitations
are collective waves. Experimental data\cite{Chang96} seem to support
this picture at filling factor $1/3$. 

In many experiments the potential confining the electrons
is slowly varying on the
scale of the magnetic length $l=(\hbar c/eB)^{1/2}$. Under these
conditions, the edge reconstructs itself\cite{Chamon94}. For
sufficiently smooth
confining potential the density profile can be determined (on a coarse
grained scale $\gtrsim l$) by the minimization of a local energy
functional\cite{Chang90_Beenaker90_Chklovskii92}. This leads to a
picture of the edge as a broad 
compressible region characterized by fractional, non--uniform
occupation of degenerate orbitals at the Fermi level. 
The single particle description is meaningless for such a state, and a
full many--body treatment of the excitations is needed. As a first step in that
direction, in a recent paper \cite{CV96} we have presented a
calculation of the spectral function of a smooth edge using a
bosonization technique inspired by previous calculations for the
uniform electron gas
\cite{JohansonKinaret94,Haussmann96,AleinerBG95_AleinerB95}. The 
theory contained a (physically motivated) cutoff on the number of
hydrodynamic modes contributing to the spectral function for a given
edge width $d$, and predicted a tunneling exponent increasing linearly
with $d$. Similar results have been obtained by Han and Thouless
\cite{HanThouless,Han97b} using a Green's function method at finite
temperature. 

In this paper we refine our calculation in such a way that short
wavelength modes are automatically excluded from contributing to the
spectral function. Thus, the present theory is {\em parameter
free}. We also extend our calculations to finite temperature
\cite{footnoteHan} and obtain the low voltage/temperature scaling of
the spectral function and of the tunneling current. Our numerical
results, in the sharp edge limit, are found to be in good agreement
with a recent experiment by Chang {\em et al.}\cite{Chang97}. 

Let us begin by writing down the microscopic Hamiltonian within the
LLL in terms of density fluctuations relative to
the equilibrium density profile $\rho_0(y)$:
\begin {equation}
H = {1 \over 2} \int_{edge} {e^2 \over \vert \vec r - \vec r' \vert}
\delta \rho 
(\vec r)  \delta \rho (\vec r') d^2r d^2r',
\label {H}
\end {equation}
where $\rho(\vec r)=\delta\rho(\vec r)+ \rho_0(y)$ is the density
operator projected  in the LLL. 
The hamiltonian (\ref{H}) is diagonal in terms of the
operators\cite{CV96} 
\begin {equation}
b_{nk} = 
{1\over  \sqrt{k l^2 \bar\rho L}} 
\int_0^L dx  e^{-ikx} \int_{0}^d dy 
f_{nk}(y)\delta \rho (x,y),
\label {deltaronk}
\end{equation}
where $f_{nk}(y)$ are the  solutions of the equation
\begin {equation}
\int_{0}^d K_0(k \vert y-y' \vert)f_{nk}(y'){\rho_0'(y')
\over\bar\rho}dy' = 
{1 \over \lambda_{nk}} f_{nk}(y),
\label {eigenvalueproblem}
\end {equation}
and $K_0(y)$ is the modified Bessel function.  
They form a complete set and satisfy the
orthonormality condition 
$\int_{0}^d f_{nk}(y)f_{mk}(y){\rho_0'(y)
\over \bar\rho}dy = \delta_{nm}$, and vanish outside the interval 
$[0,d]$.
Equation (\ref{eigenvalueproblem}) is equivalent to the 
eigenvalue problem obtained by Aleiner and Glazman \cite{Aleiner94}
from their hydrodynamical treatment of the collective modes of the
edge. 
The eigenfunction of the $n$-th branch has $n$ nodes 
in the $y$ direction and energy $\omega_{nk}=k \bar \nu
e^2/\lambda_{nk}\pi$, where 
$\bar \nu = 2 \pi l^2 \bar\rho$ is the usual filling 
factor in the bulk.
It has been shown in Ref.~\onlinecite{CV96} that the operators $b_{nk}$
satisfy boson commutation relations in the long wavelength limit
($kl\ll1$ and $n\ll d/l$).

The electronic spectral function can be obtained within 
the independent boson model (IBM)\cite{JohansonKinaret94,Mahan}, which 
describes a single localized electron
electrostatically coupled to density fluctuations. The IBM hamiltonian is
\begin {eqnarray}
H_{IBM} &&= \sum_{nk>0} \hbar \omega_{nk} b_{nk}^\dagger b^{}_{nk}
+ \nonumber\\
&& +\psi^\dagger(\vec r) \psi(\vec r)  \sum_{nk>0} M_{nk}(y)
 [b_{nk}^\dagger e^{ikx}+b^{}_{nk}e^{-ikx}],
\label{Hindependentbosons}
\end{eqnarray}
where $\psi^\dagger(\vec r)$ is the field operator that creates an
electron in the LLL coherent state (gaussian) orbital centered at
$\vec r$, which is coupled to the bosons by the matrix element
\begin {equation}
M_{nk}(y) = {2e^2 \over \lambda_{nk}}
\int dy' e^{-(y-y')/l^2} e^{-k^2l^2/4}
f_{nk}(y')\sqrt{k \bar\rho\over \pi L}\,.
\label{matrixelement}
\end {equation}
The form of the matrix elements differs from the one used in
Ref. \onlinecite{CV96} by the presence of the gaussian convolution. The main
consequence is that $M_{nk}(y)$ is found to vanish
exponentially for large $n$ and/or $k$, and therefore short wavelength
modes are automatically excluded. 
The hamiltonian
(\ref{Hindependentbosons}) can be solved by standard methods \cite{Mahan},
within the one-electron Hilbert space.  The fermionic Green's function
$G_>(y;t) = -i \langle \psi(\vec r,t)\psi^\dagger(\vec r,0)\rangle$ is
proportional to
\begin{eqnarray}
&&\exp \left\{ \sum_{nk>0} {M_{nk}^2(y) \over
\omega_{nk}^2}\left[ 
\vbox to \ifpreprintsty  0.85\baselineskip{} \else 1.2\baselineskip{} \fi
(N_{nk}+1)(e^{-i\omega_{nk}t}-1)
\right.\right. \nonumber\\
&&\hskip2cm\left.\left.
\vbox to \ifpreprintsty  0.85\baselineskip{} \else 1.2\baselineskip{} \fi
+N_{nk}(e^{i\omega_{nk}t}-1)\right]\right\},
\label{Greensfunction}
\end{eqnarray}
where $N_{nk}$ is the Bose occupation factor. 
The local density of states is given by 
the spectral function $A_>(y,\omega)$, which is defined as 
the Fourier transform of $iG_>(y,t)/2\pi$. 
From eq.~(\ref{Greensfunction}) it can be easily shown 
\cite{Minnhagen76} that
$A_>(y,\omega)$ satisfies the integral equation
\begin {equation}
\omega A_>(y,\omega) = \int_{-\infty}^\infty
{g_y(\Omega) \over 1 - e^{-\Omega/k_BT}}
A_>(y,\omega - \Omega)
d\Omega\,,   
\label {integralequation}
\end {equation}
where
\begin {equation}
g_y(\Omega) = \sum_{nk} {M_{nk}(y)^2 \over \omega_{nk}} 
\left[\delta(\Omega -\omega_{nk})+
\delta(\Omega +\omega_{nk})\right]\,.
\label {gofomega}
\end {equation}
In general, $g_y(\Omega)$ can be obtained from the imaginary part of the
density-density response function $\chi(\vec r,\vec r',\Omega)$, 
\begin {equation}
g_y(\Omega) = \!-\!
 \int d^2r' d^2r'' v(\vec r - \vec r')v(\vec r - \vec
r'') {{\rm Im} \chi(\vec r',\vec r'',\Omega)\over\Omega\pi}\,,
\label {gmicro}
\end {equation}
where $v(r)$ has the Fourier transform $v(k)= (2 \pi e^2 /k)
\exp {(- (kl)^2/4)}$. This important connection betweeen the spectral
function and the dynamical structure factor appears to be a general
property of quasiclassical collective states in the LLL. In the case
of a {\em uniform} electron liquid in the LLL equation (\ref{gmicro})
was first written down by Haussmann\cite{Haussmann96}.

If the function $g_y(\Omega)$ has a finite limit for $\Omega \to
0$, at sufficiently small $\omega$ and $T$ ($\omega,T\ll\omega_0$) the
spectral function obeys the scaling relation
\begin {equation}
A_>(y,\omega) \propto T^{g_y(0)-1}  f\left({\omega\over k_BT},
g_y(0)\right)
\end{equation}
where \cite{Han97b}
$f(x,\alpha)$ has a finite limit for $x\to0$
and behaves as $x^{\alpha-1}$  for $x\gg 1$; in particular
$f(x,2) =x/(1-e^{-x})$.

The tunneling current $I(V)$ for vertical tunneling experiments between a
smooth edge and a metallic contact satisfies a similar scaling
relation\cite{footnoteKane}, 
\begin{equation}\label{eqcurrentscal}
I(V)  \propto T^{g_y(0)}  h\left({V\over k_BT}, g_y(0)\right)\,,\end{equation}
where
\begin{equation}\label{eqcurrentint}
h(x,\alpha) =  \int f(y,\alpha) {e^{-x}-1\over e^{y-x}+1} dy\end{equation}
is linear at small $x$ and proportional to $x^\alpha$ at large
$x$. 
Therefore Ohm's law will be satisfied for $V\ll k_BT$, while
for $V$ larger than $k_BT$ but still small on the electronic energy
scale $\omega_0$ one gets a power law $I(V)\sim
V^{g_y(0)}$. The normal Fermi--liquid behaviour is recovered for
$g_y(0)=1$  ($h(x,1)=x$ at any $x$).

The calculation of the exponent $g_y(0)$ is easily
performed. Neglecting the weak nonlinearity of the  
$n=0$ mode we obtain $\alpha=g_y(0) \simeq \sum_n \beta_n(y)$, where
\begin {equation} 
\beta_n(y) = {1 \over \bar \nu}f_{n0}^2(y) e^{-n^2/2 m^2}
\label {exponent} 
\end {equation}
and $m=d \sqrt2 /\ell\pi$. The gaussian factor accounts for the
convolution with the gaussian wavefunction in equation
(\ref{matrixelement}). 

We observe that independently of the shape of the density profile
$\beta_0(y) = 1/ \bar \nu$, with negligible corrections
arising from the weak nonlinearity of the dispersion of the 
$n=0$ mode.
In the sharp edge limit, when only one branch of edge 
waves exists,
we obtain  $A_>(\omega) \sim \omega^{1/\bar \nu-1}$. For $\bar\nu=1/3$
this agrees with both CLL theory and experiment\cite{Chang96}. 
The exponent $g_y(0)\simeq d/l\bar\nu$ increases linearly with $d$ and therefore in
the limit $d \to \infty$ (limit of infinitely smooth edge) 
the tunneling density of states vanishes at low energy faster than any power 
law. This is consistent with previous results for the uniform
system\cite{JohansonKinaret94,Haussmann96}.  

\ifpreprintsty\else
\begin{figure}
\psfig{figure=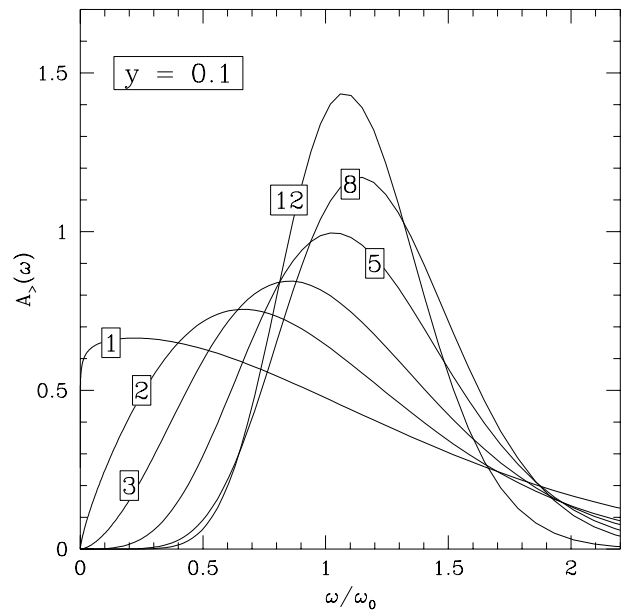,width=0.99\columnwidth}
\caption{Electronic spectral function $A_>(\omega)$ as a function of
$\omega/\omega_0$, where $\omega_0=\bar \nu e^2/\pi l$, 
for edges of a $\bar \nu=1$ QH system with $m=$ 1, 2, 3, 4, 6 and
12, at an intermediate position $y=0.6d$. 
Logarithmic corrections to the edge magnetoplasmon dispersion are
here neglected.} 
\label{figuraAw}
\end{figure}
\begin{figure}
\psfig{figure=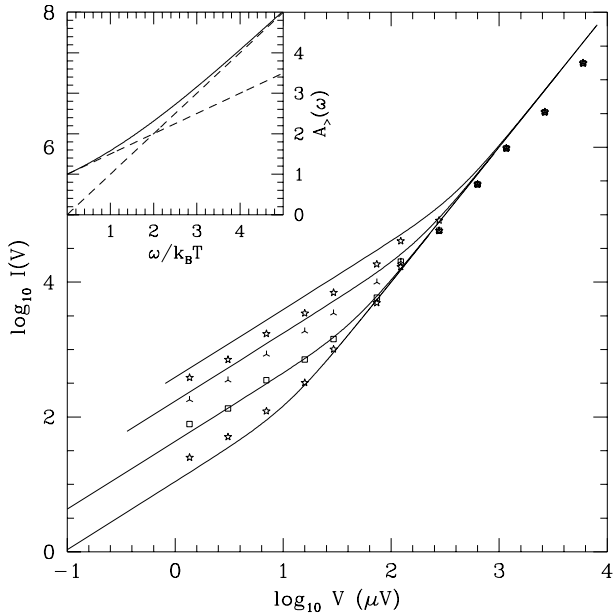,width=0.99\columnwidth}
\caption{Tunneling current $I(V)$ between a metallic contact and a
sharp edge of a $\nu=1/2$ QH fluid at various temperatures, on a
bilogarithmic scale. Full curves give the present result, points are the
experimental data by Chang {\em et al.}\protect\cite{Chang97}. The
inset shows the corresponding small--$\omega$ spectral function
$A_>(\omega)$, on a linear scale, compared with the low and high frequency
asymptotic behaviours (dashed lines).}  
\label{figChang}
\end{figure}
\fi

In Figure \ref{figuraAw} we present our numerical results for the full
electronic spectral function, calculated from 
eq.~(\ref{integralequation}) for different edge widths $d$. Notice
that the calculation is free of adjustable parameters. The eigenfunctions
$f_{nk}(y)$ and eigenfrequencies $\omega_{nk}$ have been
calculated solving numerically equation (\ref{eigenvalueproblem}) at
finite $k$ in the special case 
$\rho_0(y) = \bar\rho \left[{1\over\pi}{\mathrm{asn}}{2y-d\over d} +
{1\over2}\right]$\cite{footnoterho0}. This form of $\rho_0(y)$ allows
an analytical 
solution at small $k$, which gives
$\beta_n(y) = {1 \over \bar \nu}  T_{n}^2\left({2y-d\over
d}\right)(2-\delta_{n0}) 
e^{-n^2/2m^2}\sim e^{-n^2/2m^2}/\bar \nu$,
where $T_n(y)$ is the $n$-th Chebyschev polynomial. 

Figure \ref{figChang} compares our low--frequency results with the
experimental data by Chang {\em et al.}\cite{Chang97} for tunneling between a
metallic contact and  a nominally atomically sharp edge of a
$\nu=1/2$ QH system. The current has been obtained from equations
(\ref{eqcurrentscal}--\ref{eqcurrentint}) assuming that only
the $n=0$ charged mode contributes to the response, i.e. with an
exponent $\alpha=2$. Both the Ohmic behaviour at low voltages
 and the non--Fermi liquid behaviour at larger voltages are in good
agreement with experiment. In contrast to this, a recent composite
fermion theory by Shytov {\em et al.}\cite{Shytov97} predicts
$\alpha=3$ in this case. 

In conclusion we have obtained via a bosonization approach a
 general relation between 
 the one--electron spectral function and the dynamical structure
 factor. This allows a parameter--free calculation of the spectral
 function which interpolates between 
 the sharp edge limit and the uniform electron gas and is in good
 agreement with experiments on sharp edges both at compressible and
 incompressible filling factors.

 We gratefully acknowledge support from NSF
grant No. DMR-9706788 and from INFM.
We also acknowledge the hospitality of the Max Planck Institute 
for Physics of Complex Systems in Dresden where part of the work  has
been done.

\ifpreprintsty
\begin{figure}
\caption{Electronic spectral function $A_>(\omega)$ as a function of
$\omega/\omega_0$, where $\omega_0=\bar \nu e^2/\pi l$, 
for edges of a $\bar \nu=1$ QH system with $m=$ 1, 2, 3, 4, 6 and
12, at an intermediate position $y=0.6d$. 
Logarithmic corrections to the edge magnetoplasmon dispersion are
here neglected.} 
\label{figuraAw}
\end{figure}

\begin{figure}
\caption{Tunneling current $I(V)$ between a metallic contact and a
sharp edge of a $\nu=1/2$ QH fluid at various temperatures, on a
bilogarithmic scale. Full curves give the present result, points are the
experimental data by Chang {\em et al.}\protect\cite{Chang97}. The
inset shows the corresponding small--$\omega$ spectral function
$A_>(\omega)$, on a linear scale, compared with the low and high frequency
asymptotic behaviours (dashed lines).}  
\label{figChang}
\end{figure}

\clearpage
\psfig{figure=ce2_0.1.ps,width=0.99\columnwidth}
\vskip2cm 
\centerline{Fig. \ref{figuraAw}}
\clearpage
\psfig{figure=tfin.ps,width=0.99\columnwidth}
\vskip2cm 
\centerline{Fig. \ref{figChang}}
\fi


\begin{thebibliography}{10}
\bibitem{footnoterho0} This form is essentially equivalent to the one
chosen in \protect\cite{Aleiner94} for a gate--confined system.
\end{thebibliography}
\end{document}